# Switching mechanism of CO$_2$ by alkaline earth atoms decorated on g-B$_4$N$_3$ nanosheet

Shivam Kansara[1], Sanjeev K. Gupta[2,*], Yogesh Sonvane[1,*] and Anurag Srivastava[3]

[1]Advanced Materials Lab, Department of Applied Physics, S.V. National Institute of Technology, Surat 395007, India

[2]Computational Materials and Nanoscience Group, Department of Physics, St. Xavier's College, Ahmedabad 380009, India

[3]Advanced Materials Research Group, CNT Lab, ABV-Indian Institute of Information Technology and Management, Gwalior 474010, India

## Abstract

The adsorption and desorption of carbon dioxide (CO$_2$) molecule by alkaline earth metal (AEM) (Mg$^{+2}$, Ca$^{+2}$, Sr$^{+2}$ and Ba$^{+2}$) functionalized on graphitic boron nitride (g-B$_4$N$_3$) nanosheet have been analyzed by using density functional theory (DFT) approach includes long-range correlation (DFT+D). This method has been implemented in such a way to understand the switchable or capture/release mechanism of CO$_2$ molecule by computing the electron mobility, electronic properties, charge accumulation, charge transfer (e$^-$) and adsorption energy (Kcal/mol). The g-B$_4$N$_3$ nanosheet yields high carrier mobility ($\approx$8020 cm$^2$ V$^{-1}$s$^{-1}$) at 300 K. The positive alkaline earth adatoms on the nanosheet of g-B$_4$N$_3$ has been provided external energy to do the capture/release process of greenhouse gas CO$_2$. Here, Mg positive ion work as adatom which confirms physisorption while others show chemisorption behaviors. Therefore, due to the weak absorption of CO$_2$, it makes possible to discharge from the g-B$_4$N$_3$ nanosheet and shows instantaneous switching mechanism. Briefly, the negatively charged g-B$_4$N$_3$ nanosheets are highly sensitive for CO$_2$.

*Corresponding author(s):  yas@phy.svnit.ac.in    (Dr Yogesh Sonvane)

sanjeev.gupta@sxca.edu.in   (Dr Sanjeev K. Gupta)

# Introduction

In the last few years, the greenhouse emission has become a great concern for the whole scientific community as it affected the climate badly. As from the results, various methods have been proposed to separate, capture, store and detect greenhouse gases [1–3], with a significant contribution due to $CO_2$ in global warming [4,5]. Here, the key issues are predicting / proposing new materials as efficient to $CO_2$, capturing, storing and/or modifying materials [6,7]. Though in industries, the most common adsorbent being utilized for $CO_2$ capturing is aqueous amine solutions or chilled ammonia [8,9], the method suffers from low productivity, equipment corrosion, and solvent loss. To make this process a cost-effective and robust, the choice of adsorption material is critical. Therefore, various solid materials have been attempted as a $CO_2$ adsorbent, like metal-organic frameworks (MOFs) [10–12], porous graphene [13,14], carbon nanotube [15], silicon carbide (SiC) nanotube [16], graphene nanosheet [17] and aluminium nitride (AlN) nanosheet [18].

Recently, hexagonal boron-nitride (*h*-BN) monolayer, a structural analogue of graphene, has attracted enough attention of the scientific community. The periodic group III-V compounds, with high thermal, chemical stability, and high conductivity are suitable for different applications in electronics, optoelectronics and catalysis applications [19,20]. Due to strong covalent forces, all the atoms of each layer of BN are tightly bound together and removal of one N-atom creates defect or vacancies. It is a well-known fact in the multilayer system, where weak van der Waals (vdW) forces are responsible to hold them together as graphite. This has been the key motivation for choosing g-$B_4N_3$ nanosheet as a $CO_2$ adsorbent. Experimentally, *h*-BN films drawn from bulk BN crystal using micro-mechanical cleavage and a dielectric layer [21] with intrinsic iconicity in B-N bonds and high surface areas of *h*-BN nanosheet, as well as its tubular analogues, suggested for the gas separation or as a gas storage [22–26].

There are number of experimental investigation for $CO_2$ adsorption using zeolites and mesoporous adsorbents in the literature. A previous study of adsorbents for the $CO_2$ removal has been commonly used such as 13X zeolites [27,28], metal-organic frameworks (MOFs) [29], activated carbons [30], mesoporous silica [31,32] and surface-functionalized silica [33]. Meanwhile recently, electro-catalytically switchable $CO_2$ capture scheme has been proposed for a controlled high selective and reversible *h*-BN nanosheet [34,35]. Further, the $CO_2$ molecules are weakly adsorbed (i.e. physisorbed) on neutral *h*-BN nanosheet. In earlier studies [34–36], the $CO_2$ capture/release has been reported spontaneously and it may be merely controlled. *h*-BN is a wide-gap semiconductor with band gap around 5.8 eV [36]. Generally, the metal atoms should not aggregate to form clusters [37] thus in our study, metal atoms (Mg, Ca, Sr, Ba) have supportive substrate of g-$B_4N_3$. By adding positive ions from group II elements on g-$B_4N_3$ nanosheet, except beryllium (Be) and radium (Ra), confirms the $CO_2$ adsorption. Further, the chemically adsorbed $CO_2$ has been released and electrons are ejected. Due to the absence of the experimental and theoretical data of g-$B_4N_3$ nanosheet, the texture properties are also unknown. Meanwhile, the graphitic carbon nitride nanosheet has been successfully synthesized [38–42] and described its synthesis, functionalization, and applications process [43,44] so g-$B_4N_3$ nanosheet can be synthesized as the same.

Here, we show that an electro-catalytically switchable mechanism of $CO_2$ capture is indeed possible on a conductive g-$B_4N_3$ nanosheet; there is no reported work with same structural configuration. So we have compared our work with g-$C_4N_3$ nanosheet [45]. The present work also shows that the greenhouse gas $CO_2$ capture/release is spontaneous and improved with positive alkaline earth adatoms on the nanosheet of g-$B_4N_3$. These cycles are merely controlled and reversed without any switching on/off the charging voltage. In addition, these negatively charged g-$B_4N_3$ nanosheets with alkaline earth adatoms are good for the switching $CO_2$ from mixtures of gases in the atmosphere. This manuscript is divided into 4 sections. In upcoming section 2, the methodology has been discussed, followed by results and discussions in section 3 and conclusions in section 4.

**Methodology**

The present computational analysis of $CO_2$ capturing has been analyzed using Quantum Espresso (QE) code [46]. The generalized gradient approximation (GGA) in Perdew-Burke-Ernzerhof (PBE) [47] as well as HSE06 [48] type functionals have been customized for the treatment of the exchange-correlation. As the standard PBE functional is inefficient to define accurately weak correlation effects, the long-range interaction approach of DFT+D [49,50] has been used in the calculations. In all the computations, the substrate and adatoms have been assumed fixed, whereas $CO_2$ molecule relaxed. Further, the system is represented as the three-layer system. Where, the vacuum has been approximated as 20 Å from the upper surface as $CO_2$ of the system, which is large enough to avoid the interaction with the periodic images. In this case, 2×2 supercells with 8×8×1 Monkhorst-Pack k-point mesh [51] have been taken. The total energy includes the Kohn-sham energy, weak interaction as vdW and $E_{DFT}$ are described as $E_{disp}^{(2)}$

$$E_{DFT-D} = E_{KS-DFT} + E_{disp}^{(2)} \tag{1}$$

$E_{disp}^{(2)}$ is given by,

$$E_{disp}^{(2)} = -s_6 \sum_{i=1}^{N_{at}-1} \sum_{j=i+1}^{N_{at}} \frac{C_{ij}^6}{R_{ij}^6} f_{dmp}(R_{ij}) \tag{2}$$

Where $N_{at}$ represents the number of present atoms, $C_{ij}^6$ shows the dispersion coefficient, $R_{ij}$ signify the distance between atoms for atom pair $ij$, and $S_6$ as a scaling factor, subject to the exchange correlation (XC)-functional. A damping factor $f_{dmp}$ is utilized to inhibit singularities at few distances.

To investigate the adsorption on adsorbent, the adsorption energy ($E_{ads}$) of $CO_2$ molecule on g-$B_4N_3$ can be calculated as below [45]

$$E_{ads} = \frac{(E_{complex} - E_{(B_4N_3+adatoms)} - E_{CO_2})}{n} \tag{3}$$

Where $E_{complex}$, $E_{(B_4N_3+adatoms)}$ and $E_{CO_2}$ are the total energy of a complex system, the energy of nanosheet with adatoms, and energy of an adsorbed $CO_2$ molecule on g-$B_4N_3$, respectively.

## Results and discussions

### a. Structural and electronic properties

Fig. 1(a) and (b) shows the lowest energy configuration of structure and electron mobility of isolated g-$B_4N_3$ nanosheet. The unit cell of g-$B_4N_3$ nanosheet consists of seven atoms, where, four atoms are of B and three atoms of N. The average bond length between B-B, B-N and B-B is 1.43 Å with the buckled angle of 113º. In case of the static structure of g-$B_4N_3$, the value of electron mobility is ≈8020 cm$^2$/V·s for 300 K and computed using ab initio model for mobility and Seebeck coefficient using Boltzmann transport (MoBT) equation [52]. High electron mobility is a prerequisite for injecting extra electrons into electro-catalytically switchable $CO_2$ capture materials. This readily facilitates the electron injection/release for electrocatalytically switchable $CO_2$ capture. The value of electron mobility of planer *h*-BN monolayer is between the range of 9200-28800 cm$^2$/V·s [53], which is slightly larger than presenting monolayer.

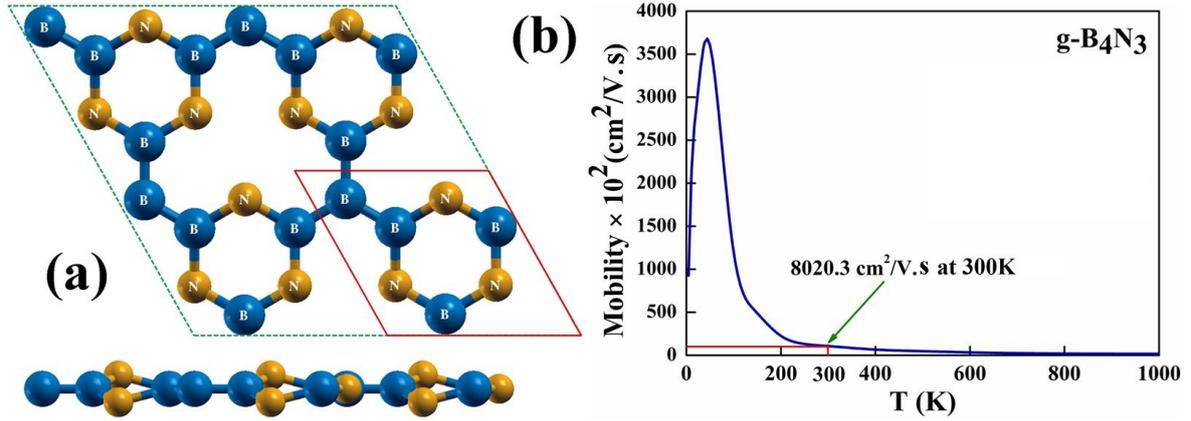

**Figure 1:** Top (upper) and side (lower) display **(a)** nanosheet of g-$B_4N_3$ where the defect has been created by removing one nitrogen atom from unit cell. The blue and yellow balls represent B and N atoms, respectively, and the unit cell of g-$B_4N_3$ is represented by a dark red line. The calculated electron mobility **(b)** of nanosheet of g-$B_4N_3$ with varying temperature and the red line represents the mobility at 300K temperature.

**Table 1**: Calculated formation energy using electronic energy and its comparison with earlier reports for different 2D material per atom.

|  | **Structure** | **Formation energy (eV)** |
|---|---|---|
| Reported work [39,40] | graphene | -84.95 |
|  | silicene | -38.20 |
|  | $Si_2BN$ | -50.38 |
|  | BN | -81.70 |
| Present work | g-$B_4N_3$ | -46.76 |

At ≈50 K, the electron mobility is at its highest peak ≈ 3635 ×$10^2$ cm$^2$/V·S. The electron mobility can be defined by the equation (4);

$$\upsilon_d = \mu E \qquad (4)$$

Where, $\upsilon_d$ is the drift velocity; $\mu$ and $E$ is electron mobility and magnitude of the electric field, respectively. Table 1 represents the formation energy for the different system at 0 K. These energies were calculated for the unit cell and the formation energy of g-$B_4N_3$ has a good comparison to the $Si_2BN$ [54]. The previously reported works [55,56] were also calculated the formation energy using chemical potential instead of electronic energy of defected h-BN nanosheet. The high formation energy of g-$B_4N_3$ confirms the stability of the structure. The formation energy ($E_f$) using chemical potential of doped AEM sheets have been computed through the following equation (5);

$$E_f = E_{defect} - E_{pristine} + n_B\mu_B + n_N\mu_N - n_d\mu_d \tag{5}$$

In equation 5, $E_{defect}$ and $E_{pristine}$ are the total energies of considered defect sheet and pure sheet of boron nitride (BN). $\mu_B$, $\mu_N$ and $\mu_d$ are chemical potentials of boron, nitrogen and alkaline earth atoms, respectively. n represents the number of atoms present in the sheet. The values of the formation energy are shown in the ESI Table S2.

The band structure and projected density of states (PDOS) have been computed for the nanosheet of g-$B_4N_3$ and is shown in Fig. 2(a) and (b) using the PBE functional. The electronic band profile has been calculated with respect to the highest symmetries in the first Brillouin zone for an energy range of -6 to 6 eV. In the band curve $\Gamma$- $M$- $K$ plane of the Brillouin zone (BZ) bands split, due to the defect introduced by replacing one atom of N from the unit cell. It has been assumed that by replacing one N atom from the $h$-BN nanosheet, the Fermi level gets shifted downward to the valence band. From band structure Fig. 2 (a), the nanosheet of g-$B_4N_3$ is found metallic. The calculated band structure and PDOS using HSE06 functional are shown in Electronic Supplementary Part (ESI) as Fig. (S1).

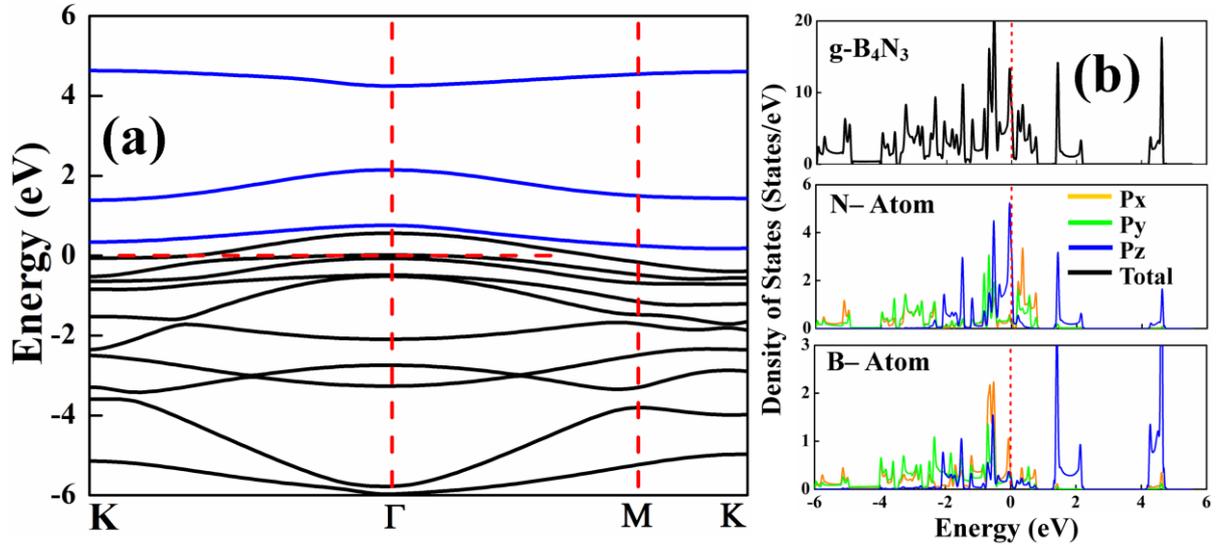

**Figure 2:** The calculated **(a)** electronic band curves of nanosheet of g-$B_4N_3$. The blue and black lines show the conduction and valence band, respectively. **(b)** Density of states of a nanosheet of g-$B_4N_3$. The red dotted line represents the Fermi level.

To understand the electronic coupling and orbital contributions, the PDOS profile has also been computed, shows the contribution of subshell from the last orbital (*p*) of B and N atoms, in the states at Fermi level. The $2p_x$ subshell and outermost $2p_z$ subshell of B and N atom are much contributed at the Fermi level where N atom is more dominant to the B atom. For the g-$B_4N_3$ nanosheet, the total PDOS shows the mixed character of the B and N atoms of their conduction band (CB) and valence band (VB) channels. VB is more dominated by B-$2p$ orbitals in comparison to small contribution from N-$2p$ orbitals. Furthermore, B-$2p$ orbitals dominate at the CB, while all the details regarding adsorption energy and bader charge have been illustrated in Fig. 4. Figure 3 (a-d) shows the partial density of states of the AEM doped BN sheets. The $p$ orbital of B and N are highly contributed near and at the Fermi level in Figure 3. As we can see that as increase the number of electron of the systems, $s$ and $p$ orbital of alkaline metal and $CO_2$ molecule is going to be more dominate near the Fermi level, respectively. This hybridization affects the bonding of the $CO_2$ molecule to the sheet at the lowest distance. The grey shaded region represents the total density of states.

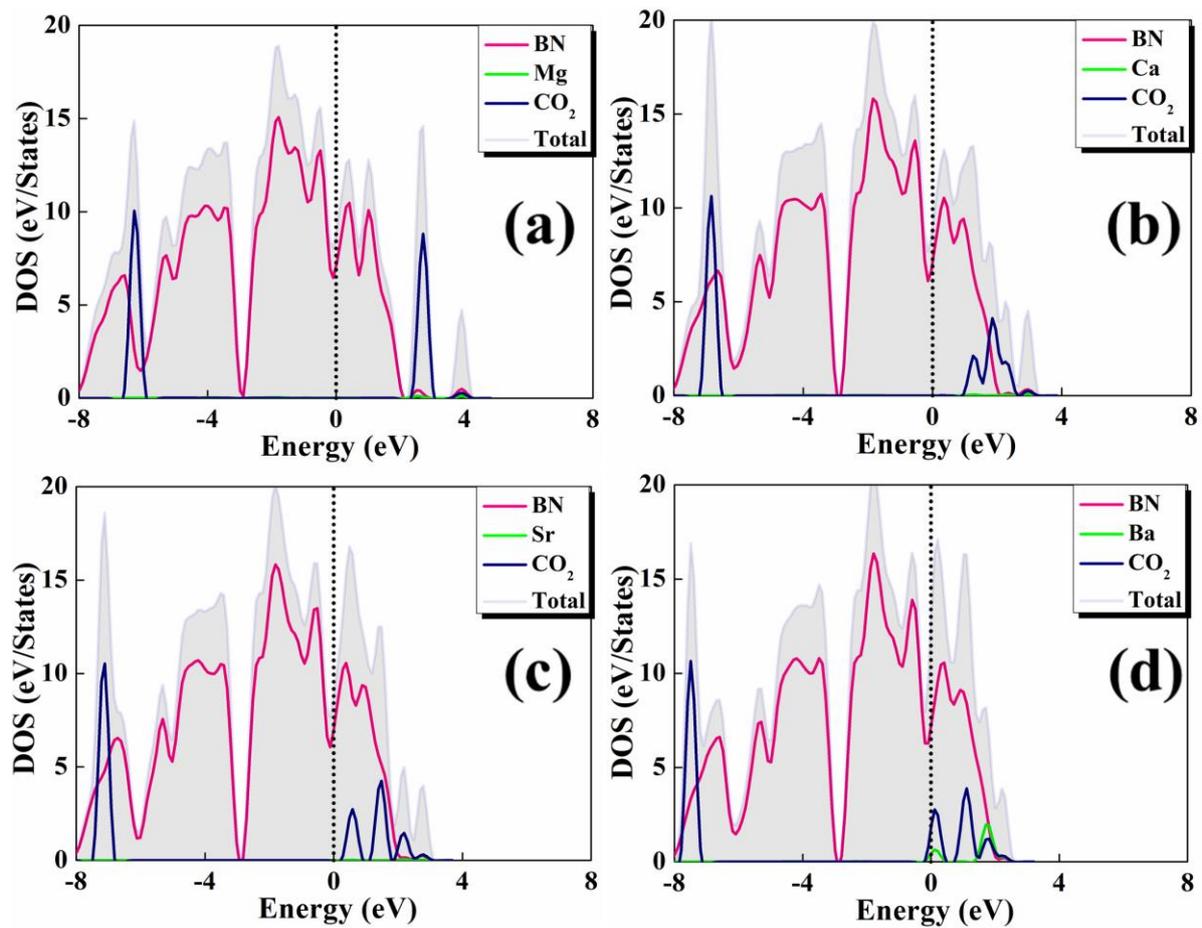

**Figure 3**: Partially density of states of group II atoms doped BN sheets. The projections of B with N and C with O atoms are considered together for the p orbitals, respectively. The dotted dashed lines represent the Fermi level.

The adsorption energies and charge transfer of $CO_2$ have been calculated on neutral and negatively charged g-$B_4N_3$ and also by adsorbing positive ions from alkaline earth elements. Fig. 4 depicts the comparative adsorption energies of $CO_2$ on g-$B_4N_3$ with alkaline earth adatoms and the charge transfer with respect to the bond length between the oxygen atoms of $CO_2$ to g-$B_4N_3$. In contrast, $CO_2$ is weakly adsorbed on the $Sr^{+2}$ atom with smallest adsorption energy. It has been noticed that reaction enhances from -0.35 kcal/mol to 9.66 kcal/mol in the case of Ba and Mg adatoms, the processes go from chemisorption to physisorption, respectively. $CO_2$ is tightly chemisorbed on g-$B_4N_3$ for the $Ca^{+2}$ adatom with large adsorption energy as -13.37 kcal/mol. The total charge transfer for $Ca^{+2}$ adatom is 0.46 |e|. The details of adsorption and Bader charge transfer for the different adatoms are shown in Fig. 3. The adsorption energy of adatoms as $Mg^{+2}$, $Ca^{+2}$, $Sr^{+2}$ and $Ba^{+2}$ with the substrate of g-$B_4N_3$ nanosheet is shown in Electronic Supplementary Part (ESI) as Table (S1).

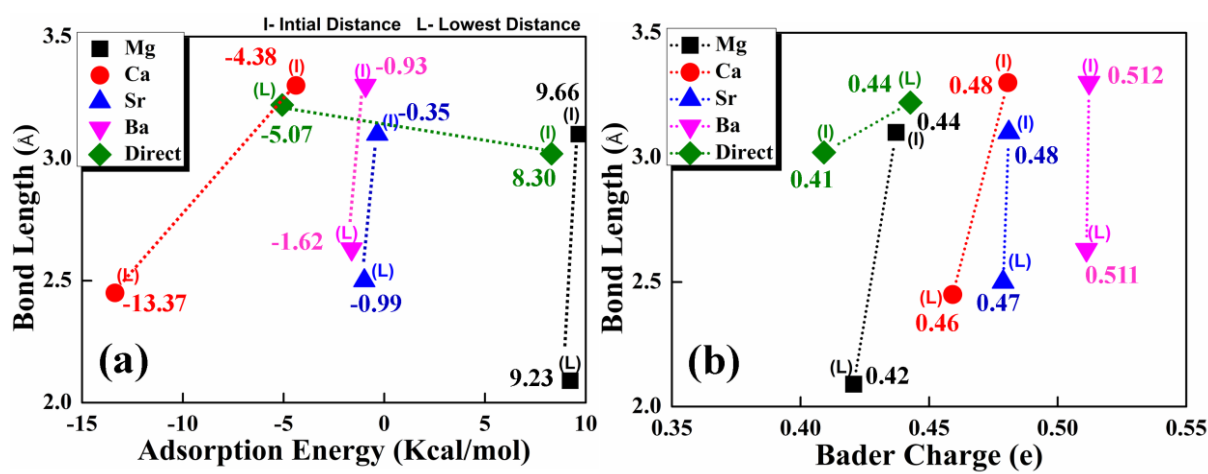

**Figure 4:** Comparison between the **(a)** adsorption energy (Kcal/mol) ($\Delta E_{DFT}$) **(b)** Charge transfer (e⁻) calculated with respect to the bond length (Å) using DFT and energies estimated using universal scaling relations ($\Delta E_{predicted}$).

## b. Adsorption of $CO_2$ molecule on the nanosheet and charge accumulation

To visualize the charge redistribution on $CO_2$ molecule, the charge density difference or charge accumulation has been computed using following equation (6);

$$\Delta \rho = \rho_{g.B_4N_3+adatom+CO_2} - \rho_{g.B_4N_3} - \rho_{adatom} - \rho_{CO_2} \qquad (6)$$

Where $\rho_{(g.B_4N_3+adatom+CO_2)}$, $\rho_{g.B_4N_3}$, $\rho_{adatom}$, and $\rho_{CO_2}$ are the total charge density of complex, density of g-$B_4N_3$, density of extra injected positive ions as adatoms and density of molecules, respectively.

Figs. **5(a-e)** addresses the adsorption process with different adatoms interacting with a negative negatively charged nanosheet g-$B_4N_3$, where distance and angle of $CO_2$ molecule have been changed due to the effect of the system. The charge transfer of total five systems

such as direct $CO_2$ molecule adsorption on nanosheet g-$B_4N_3$ along with different alkaline adatoms as $Mg^{+2}$, $Ca^{+2}$, $Sr^{+2}$, $Ba^{+2}$ atoms have been analyzed. From the Figs. **5(a-e)**, we can see that the charge transfer regions are between two interfaces of $CO_2$ molecule and nanosheet. This process is analyzed by surface hybridization of the top layer to the bottom layer, which is indicating the effect of weak vdW interlayer interaction and covalent forces. The interaction between a $CO_2$ molecule and alkaline earth adatoms exerts a driving force, these plays a basic role in the electron transfer from the molecule to the adatoms region, while holes move in the converse direction. The charge deficient region between the O=C=O concludes the loss of charge, which decreases from double bond to the O=C-O bond formation. So, the Bader charge analyzer helps to calculate the electron transfer.

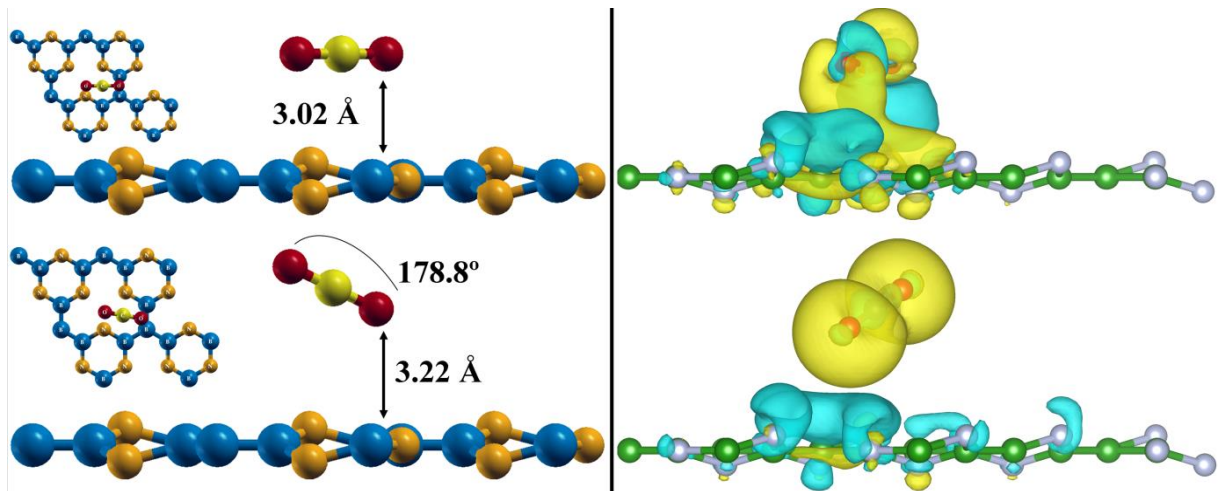

**Figure 5:** Top and side views of $CO_2$ sensing and charge accumulation on a 2×2 g-$B_4N_3$ nanosheet. Atom color code: blue and yellow atomic colors are represented as green and silver in accumulation, respectively.

Fig. **5** depicts the process of direct $CO_2$ molecule adsorption on g-$B_4N_3$ nanosheet and noticed that the distance of $CO_2$ molecule and angle have been changed as compared to initial position. Due to the strong repulsive force arising from electron density as well as a rapid increase of ground energy, the molecular relative distance has been increased. The yellow and blue colors are in the figures representing the positive and negative value of iso-surface, respectively. In all the figures, the yellow surfaces are bound to state, where the plotted amount is more prominent than a (positive) threshold, and the blue surfaces bound regions where the plotted amount is not much as an alternate (negative) threshold. The blue cloud means losing charges and yellow cloud represents the gaining of charges, respectively. At the initial structure, there is a high electron density cover the nanosheet and mixed electron density of both while after moving far from $CO_2$ gas from the substrate the electron density also get separated as shown in Fig. **5**.

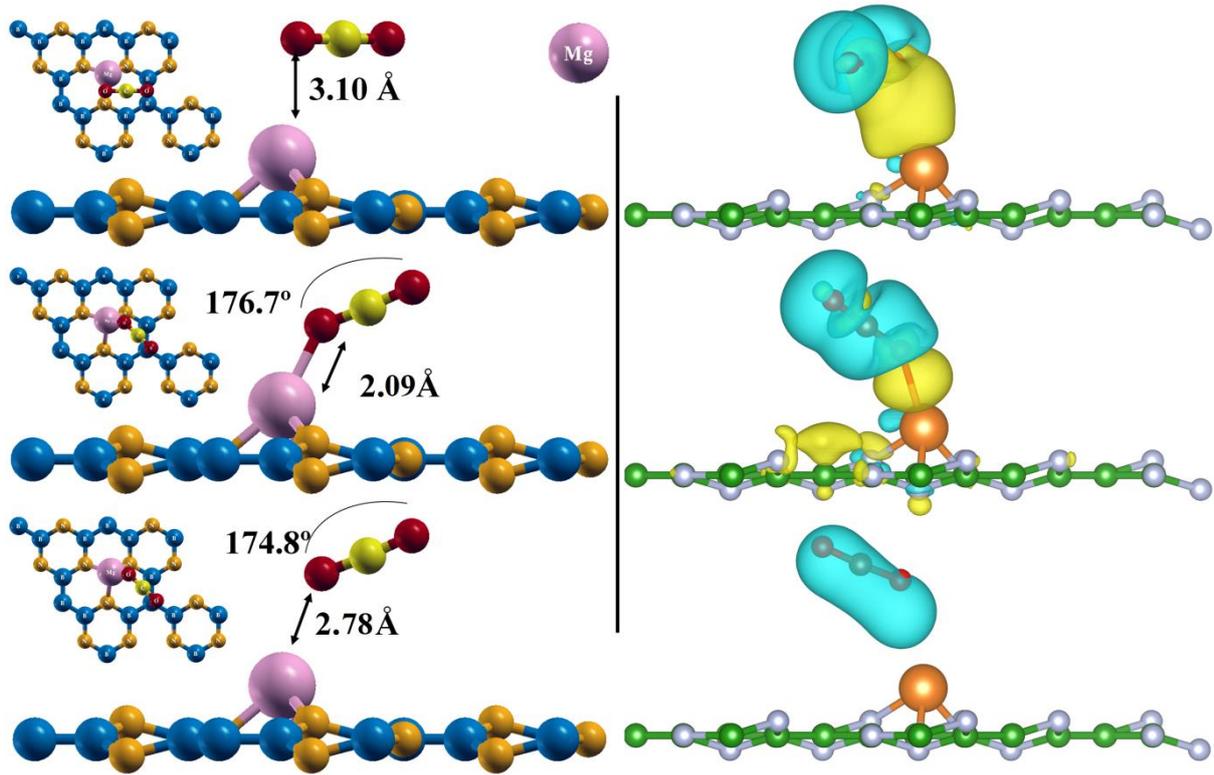

**Figure 6:** Top and side views of switching of $CO_2$ and charge accumulation on a 2×2 g-$B_4N_3$+ Mg as adatoms on nanosheet. Atom color code: blue, yellow and magenta colors as green, silver, and orange in accumulation, respectively.

Fig. **6** shows that the $Mg^{2+}$ ion makes high interaction among the molecules. From the initial state to final state, g-$B_4N_3$ nanosheet plays a significant role to make a possible switching of $CO_2$. The external injected positive ion as $Mg^{2+}$ strongly interacts with the electron enriched O atom of $CO_2$ molecule. Bader charge analysis disclosed noteworthy charge transfer (0.42 |e|) from $CO_2$ molecule to the substrate at the time of bonding. The transferred charges were non-uniformly distributed during the whole process. The reason for this interaction is the quantum mechanical change of momentary dipoles, which begins from the shared electrons in the adsorbate and the surface. However when an atom comes closer to the surface, kinetic energy of the electrons increases, as an effect of overlapping of wave functions of the atom. This is responsible for the high repulsion potential [57]. As well as also introduced the pressure dependent adsorption process in the Table S3. In this case, the adsorption energy is positive so for an easy physisorption process, the relation between the adsorbate concentration and the gas molecule can be represented by the Langmuir adsorption isotherm equation (7):

$$\theta = \frac{bp}{1+bp} \qquad (7)$$

Where, $\theta$ is surface coverage, b and $p$ are the temperature constants.

Fig. **7** shows a $CO_2$ molecule adsorbed on a $Ca^{2+}$ adatom. In this case, the adsorption energy [$E_{ads}$] increases from that for the single support of g-$B_4N_3$ nanosheet. Therefore, the supported $Ca^{2+}$ adatom would prefer to be adsorbed on O atoms from the $CO_2$ molecule. The bader charge analysis data indicate that the complex system and $CO_2$ molecule have 0.460|e| and 0.125|e| charges at the process of adsorption. And also carry a little negative charge on the adatoms. The accumulation and depletion charges are raised on the system at the time of switching, while the depletion process begins on behalf of accumulation. To understand the bonding property of $CO_2$ molecule on adatom, the two simple processes such as bonding and after repel $CO_2$ molecule like switching process is also analyzed here. The charge redistribution form corresponds to each new system, so the visual diagram could be more clearly expressed by the charge distribution. From the strong charge distribution on the accumulation or depletion recognized the strength of the bonding and distributed charge cloud determined the orbit bonding [58]. In all figures of charge transfer, blue and red color represents the charge accumulation and depletion, respectively. Here, the adsorption energy is chemisorption so that it should be chemical bonding.

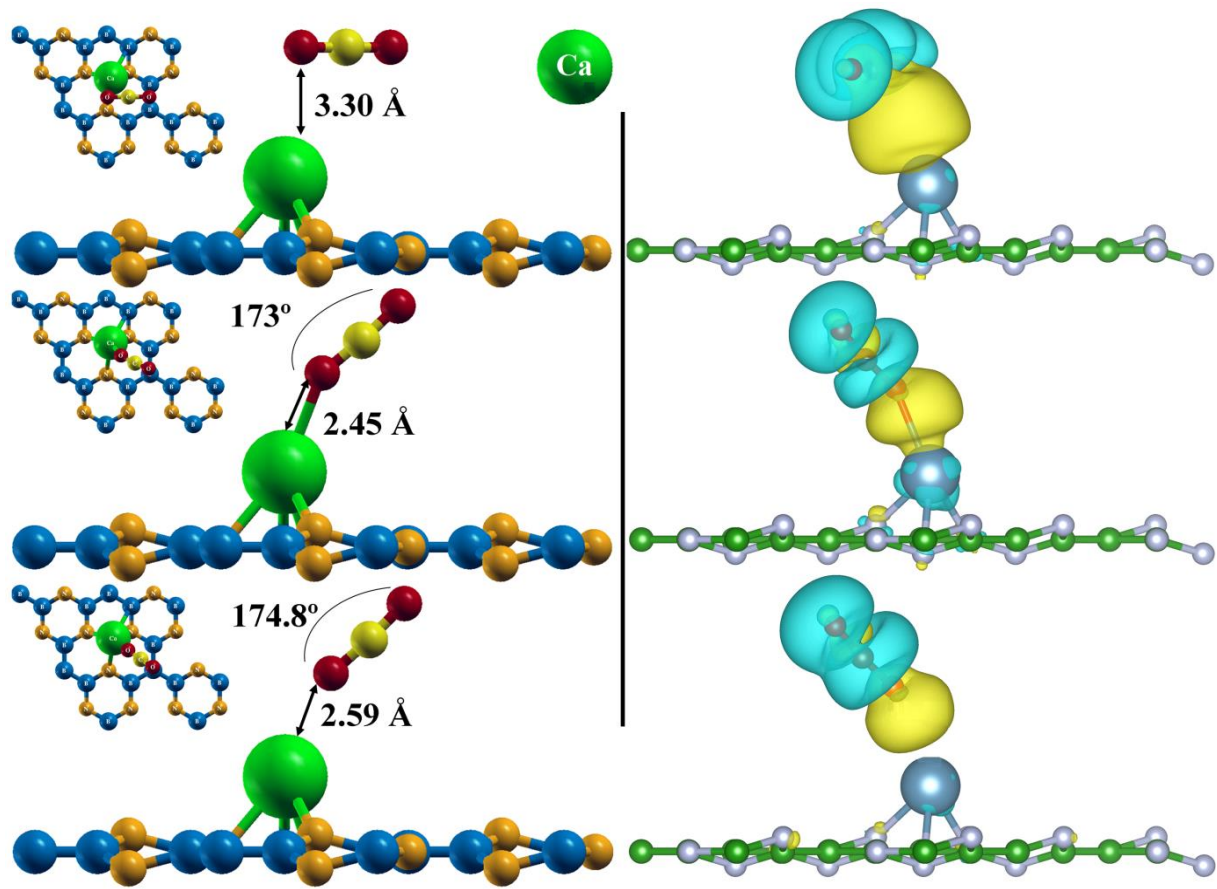

**Figure 7:** Top and side views of switching of $CO_2$ and charge accumulation on a 2×2 g-$B_4N_3$+ Ca as adatoms on nanosheet. Atom color code: blue, yellow and green colors as green, silver, and black in accumulation, respectively.

The adsorption of $CO_2$ molecule on g-$B_4N_3$ nanosheet supported with extra adatom as a catalyst $Sr^{2+}$, there are three cases displayed in this case, such as initial, lowest distance and repelled position of $CO_2$ molecule (See Fig. 8). The adsorption of the $CO_2$ molecule in the complex system or the distance of the molecule from the substrate is 3.10 Å, 2.50 Å and it repels at 2.88 Å at switching process. It is weakly chemically adsorbed on the surface with the adsorption energy of -0.99 kcal/mol. Therefore, the $CO_2$ is adsorbed on the supported catalyst as adatom $Sr^{2+}$ on the g-$B_4N_3$ support.

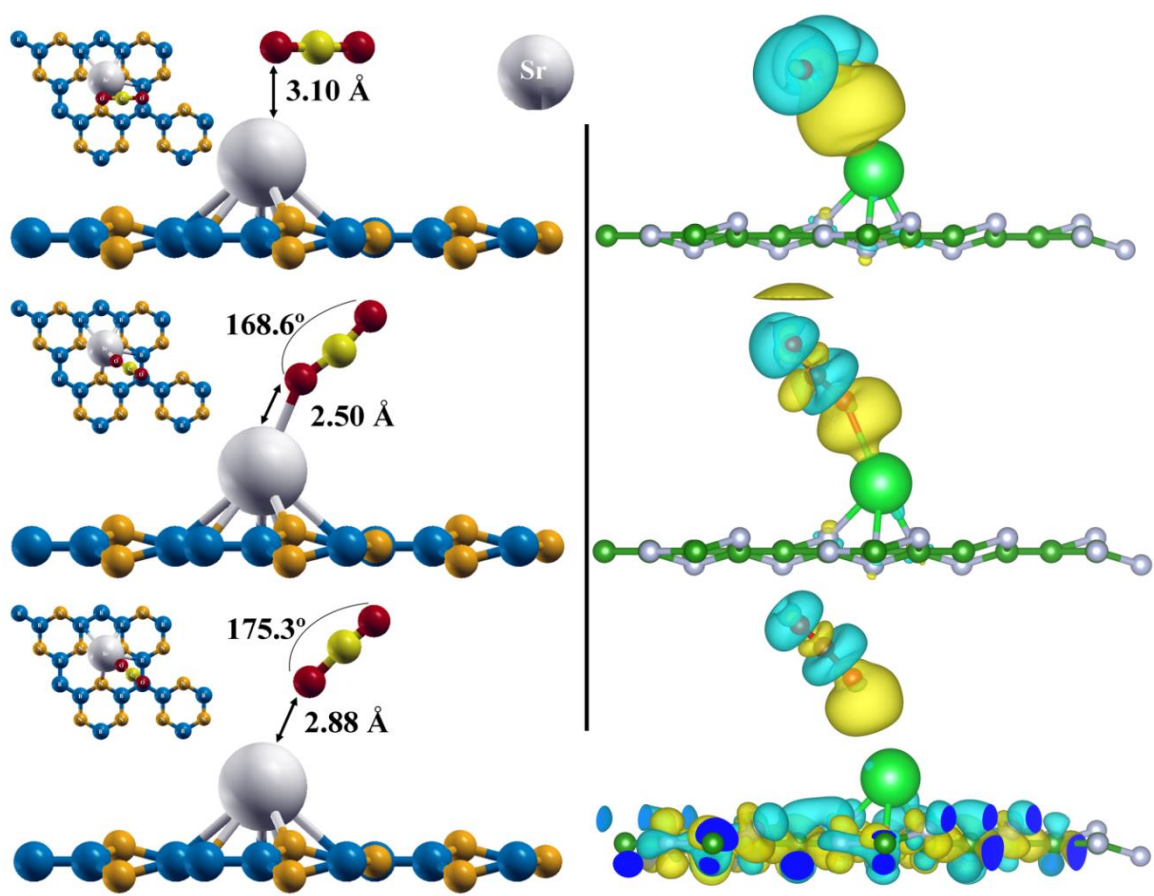

**Figure 8:** Top and side views of switching of $CO_2$ and charge accumulation on a 2×2 g-$B_4N_3$+ Sr as adatoms on nanosheet. Atom color code: blue, yellow and magenta colors as green, silver, and orange in accumulation, respectively.

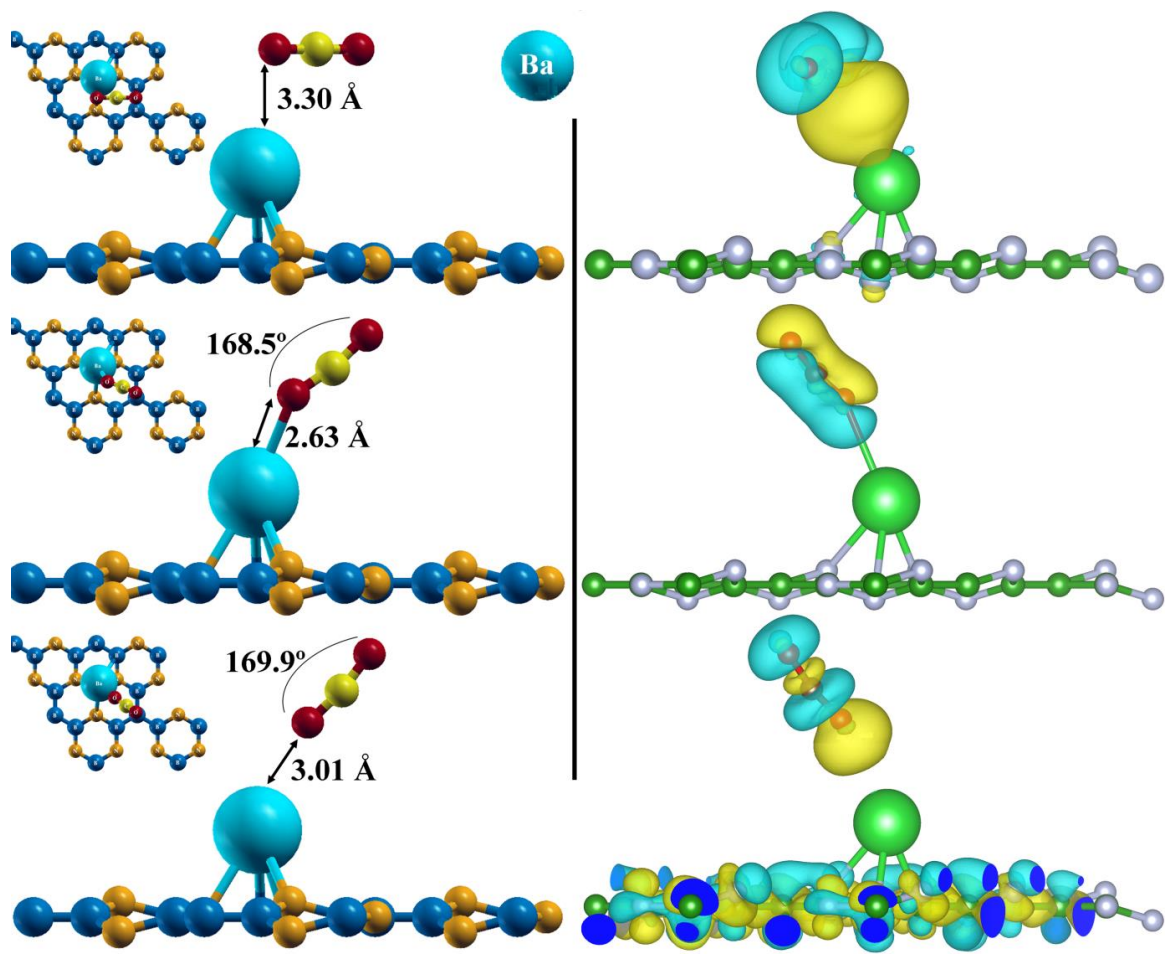

**Figure 9:** Top and side view of switching of $CO_2$ and charge accumulation on g-$B_4N_3$+ Ba as adatoms on nanosheet. Atom color code: blue, yellow and sky blue colors as green, silver, and green in accumulation, respectively.

To further explain the charge transfer phenomenon between gas adsorbates and substrate, Fig. **9** shows the iso-surface plot of electron charge density gas molecules. Here, the charge difference is calculated in three ways by subtracting the electron densities of non-interacting components (gas) from the charge density of the adatoms and initial system with the same atomic positions of the systems. It is also seen that there is a charge accumulation on the substrate for $CO_2$ adsorbate characteristic. In this case, $CO_2$ acts as an electron acceptor from the Bader charge calculation. The $CO_2$ molecule accepts the electron from the adatom as a catalyst and shows the charge depletion region between the inner surface of the molecule and the substrate. The adsorption energy of this system is -0.93 Kcal/mol, and the distance between molecules to the substrate is 3.30 Å and after making bond, it is 2.63 Å with -1.62 Kcal/mol adsorption energy. The small value of adsorption energy with respect to atomic distance indicates a weak interaction. The charge transfer between $CO_2$ to the substrate has been obtained from bader charge analysis (Fig. 4). The adsorption of $CO_2$ on $Ba^{2+}$, the calculated charges on the $CO_2$ and complex system are 0.13 |e| and 0.512 |e|, respectively. Meanwhile, very dense charges have been transferred from the upper two layers as $CO_2$ and $Ba^{2+}$. Our reported work is being used for the instant switching mechanism of the $CO_2$ molecule due to the small adsorption value with compared with other works [31] and shown in Table 2.

Table 2: The comparisons of adsorption energy with previous works have been compared.

| Work | Systems | Supportive | Adsorption Energy (Kcal/mol) |
|---|---|---|---|
| *Our* | *Graphitic BN nanosheet* | Mg | 9.23 |
| | | Ca | -13.37 |
| | | Sr | -0.99 |
| | | Ba | -1.62 |
| *Ref.* [35] | *BN nanosheet* | 0 e- | -49.57 |
| | | 1 e- | -13.87 |
| | *BN nanotube* | 0 e- | -69.21 |
| | | 1 e- | -11.91 |
| *Ref.* [59] | *C$_2$N sheet* | - | -8.05 |
| *Ref.* [44] | *Hybrid BN/BN/G sheet* | 0 e- | -5.30 |
| | | 4 e- | -59.49 |
| | *Hybrid BN/G sheet* | 0 e- | 90.16 |
| | | 4 e- | -39.43 |

## Conclusions

In summary, we have used DFT-D to investigate the adsorption and desorption of $CO_2$ molecule on g-$B_4N_3$. The structure of g-$B_4N_3$ nanosheet has high electron mobility ≈8020 cm$^2$/V·S at 300 K and the formation energy -6.67 eV/atom, which confirms the stability of the system. As well as, we also focused on the stability of the alkaline doped nanosheet. The calculations show that the switching process of greenhouse $CO_2$ is varied by using different alkaline adatoms on the surface of g-$B_4N_3$ nanosheet. Furthermore, $CO_2$ spontaneously releases electrons from the absorbent without any reaction barrier. Our results suggest that alkaline earth adatoms on g-$B_4N_3$ nanosheet are good adsorbents for $CO_2$ and they can be used to switch $CO_2$ from the atmosphere. $Mg^{+2}$ adatom takes physisorption reaction (9.23 kcal/mol to 9.66 kcal/mol) while expecting all behave as chemisorption reaction. The direct $CO_2$ molecule has been repelled from the substrate of g-$B_4N_3$. The chemisorption reaction has been enhanced from -0.93 kcal/mol to -13.37 kcal/mol for $Ba^{+2}$ to $Ca^{+2}$ adatoms. Due to the lower adsorption values of $CO_2$ to the substrate, $CO_2$ molecule makes easy to discharge from the g-$B_4N_3$ nanosheet without any external source and shows instant switching mechanism. This work is interesting one for computational predictions, which might take place an experiment possible high-selectivity mechanism of $CO_2$ switching on g-$B_4N_3$ material with the ideal reversible process.

**ACKNOWLEDGEMENTS**

Y.A.S & S. K. G. acknowledges the use of high performance computing clusters at IUAC, New Delhi and YUVA, PARAM II, Pune to obtain the partial results presented in this paper.

# Electronic Supplementary Information (ESI)

**Electronic properties:**

The calculated electronic properties for the sheet of g-$B_4N_3$ using HSE06 functional has been described in Fig. (S1).

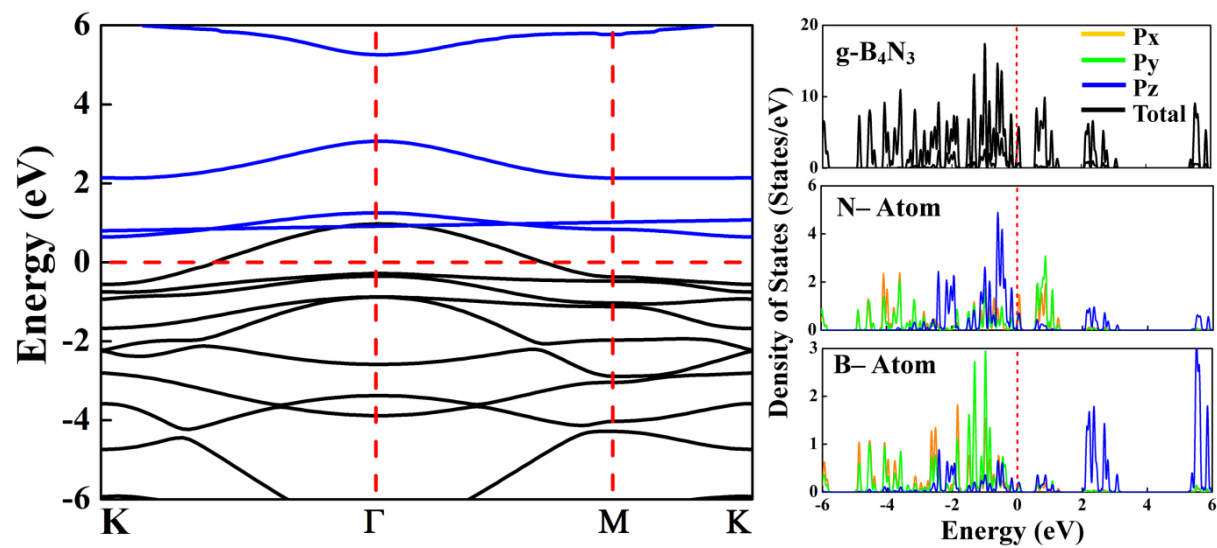

Figure S1. The calculated electronic band curve of **(a)** sheet of g-$B_4N_3$. The red dashed line represents the Fermi level. The blue and black lines in (a) show the conduction and valence band, respectively. **(b)** Density of states of a sheet of g-$B_4N_3$. The red dotted line represents the Fermi level.

**Adsorption Energy:**

**Table S1.** The adsorption energy of adatoms is as $Mg^{+2}$, $Ca^{+2}$, $Sr^{+2}$ and $Ba^{+2}$ with the substrate of g-$B_4N_3$.

| Adatoms | Adsorption Energy (Kcal/mol) with g-$B_4N_3$ |
|---|---|
| **Mg** | -111.613 |
| Ca | -140.439 |
| **Sr** | -144.128 |
| Ba | -135.596 |

**Table S2**. Represent the formation energy (eV) and comparison of adsorption energy of the C atom and the $CO_2$ molecule with the substrate of g-$B_4N_3$ along different adatoms (Mg, Ca, Sr, Ba).

| Systems / Adsorption Sites | Formation energy ($E_f$) (eV) | C- Atom (Kcal/mol) | $CO_2$- Molecule (Kcal/mol) |
|---|---|---|---|
| g-$B_4N_3$+Mg+$CO_2$ | -38.46 | -29.37 | 9.23 |
| g-$B_4N_3$+Ca+$CO_2$ | -35.53 | -35.28 | -13.37 |
| g-$B_4N_3$+Sr+$CO_2$ | -40.47 | -35.97 | -0.99 |
| g-$B_4N_3$+Ba+$CO_2$ | -40.46 | -33.89 | -1.62 |